\documentclass{article} 
\usepackage{iclr2027_conference,times}

\usepackage{amsmath,amsfonts,bm}

\def\eqref#1{equation~\ref{#1}}

\def\1{\bm{1}}

\DeclareMathAlphabet{\mathsfit}{\encodingdefault}{\sfdefault}{m}{sl}
\SetMathAlphabet{\mathsfit}{bold}{\encodingdefault}{\sfdefault}{bx}{n}

\usepackage{amsmath}
\usepackage{amssymb}
\usepackage{graphicx}
\usepackage{booktabs}
\usepackage{multirow}
\usepackage{hyperref}
\usepackage{url}

\title{History-Conditioned Joint-Prefix Alignment for Generative Recommendation}

\author{Hongliang Sun, Lianjie Li, Bolin Zhang, Dianbo Sui, Dianhui Chu \& Zhiying Tu \thanks{Corresponding author} \\
Department of Computer Science, Weihai \& Qingdao Research Institute\\
Harbin Institute of Technology\\
Weihai, China \\
\texttt{sunh@hit.edu.cn, 24s030110@stu.hit.edu.cn,} \\
\texttt{\{bolin, suidianbo, chudh, tzy\_hit\}@hit.edu.cn}
\\}

\iclrfinalcopy 
\begin{document}

\maketitle

\begin{abstract}
Generative recommendation retrieves items by autoregressively generating semantic identifiers, but beam search may discard a target before its complete identifier is generated. Our preliminary analysis across three benchmarks shows that most missed targets are pruned within the first two decoding steps, highlighting the importance of early prefix retention. However, retaining a target's first-token branch alone is insufficient if its continuation is pruned at the next step: aligning only the first-token distribution improves first-token survival but yields little gain in complete-path retention. This observation motivates joint supervision of early branches and their continuations.
We propose \textbf{P}refix \textbf{A}lignment with \textbf{T}emporal \textbf{H}istory (PATH), which aggregates transition statistics from the training corpus over recent interactions with exponential decay to construct history-conditioned two-token prefix targets. PATH aligns the model's joint predictions with these targets through forward KL divergence, using a chain-rule decomposition that enables unbiased Monte Carlo estimation. At inference time, PATH reuses the transition statistics for pointwise mutual information (PMI) calibration, reranking completed candidates relative to global prefix frequency. Experiments on Beauty, Instruments, and Yelp demonstrate the effectiveness of PATH, showing consistent improvements in recommendation performance and higher full-SID survival rates.
\end{abstract}

\section{Introduction}
\label{sec:intro}

\begin{figure}[t]
\centering
\begin{minipage}[t]{0.515\linewidth}
\vspace{0pt}
\centering
\includegraphics[width=\linewidth,trim=18 76 48 14,clip]
{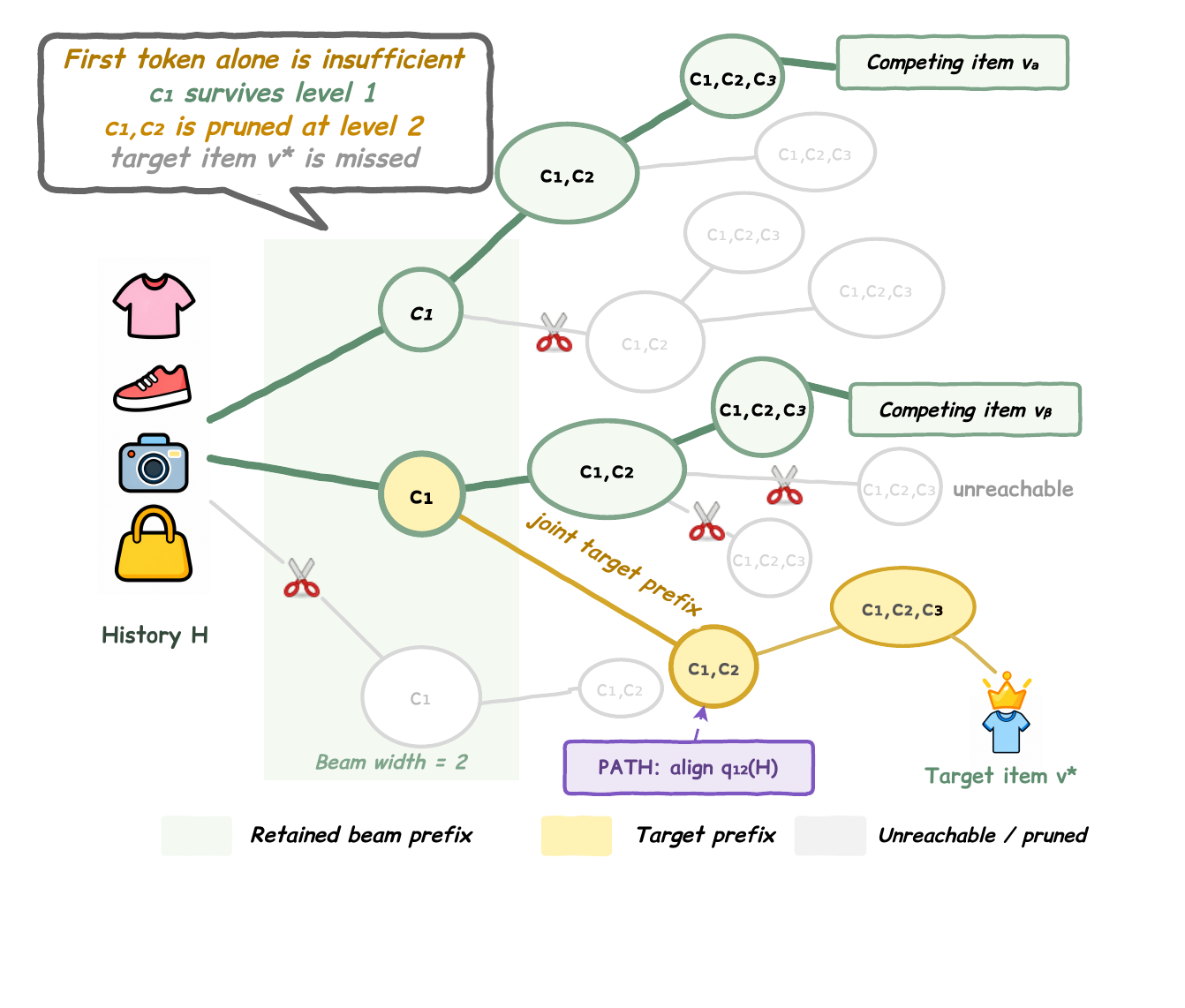}
\end{minipage}\hfill
\begin{minipage}[t]{0.470\linewidth}
\vspace{0pt}
\centering
\begin{minipage}[t]{0.485\linewidth}
\vspace{0pt}
\centering
\includegraphics[width=\linewidth]
{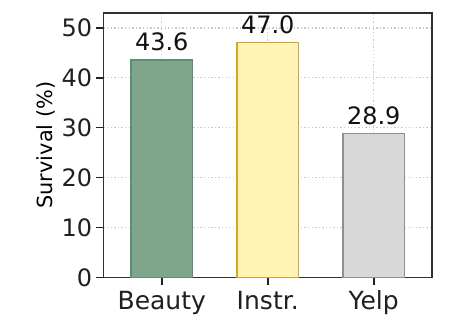}
\par\vspace{-1.5mm}
\makebox[\linewidth][l]{\hspace*{0.23\linewidth}\makebox[0.73\linewidth][c]{\scriptsize After $c_1$}}
\end{minipage}\hfill
\begin{minipage}[t]{0.485\linewidth}
\vspace{0pt}
\centering
\includegraphics[width=\linewidth]
{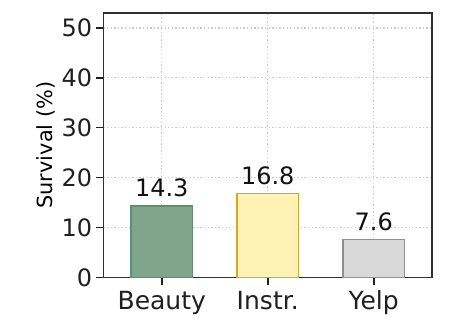}
\par\vspace{-1.5mm}
\makebox[\linewidth][l]{\hspace*{0.23\linewidth}\makebox[0.73\linewidth][c]{\scriptsize After $c_2$}}
\end{minipage}

\par\vspace{3.0mm}
\begin{minipage}[t]{0.485\linewidth}
\vspace{0pt}
\centering
\includegraphics[width=\linewidth]
{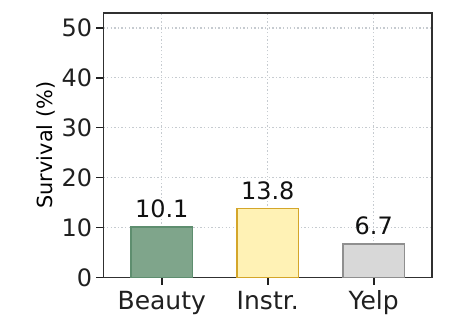}
\par\vspace{-1.5mm}
\makebox[\linewidth][l]{\hspace*{0.23\linewidth}\makebox[0.73\linewidth][c]{\scriptsize Full SID}}
\end{minipage}\hfill
\begin{minipage}[t]{0.485\linewidth}
\vspace{0pt}
\centering
\includegraphics[width=\linewidth]
{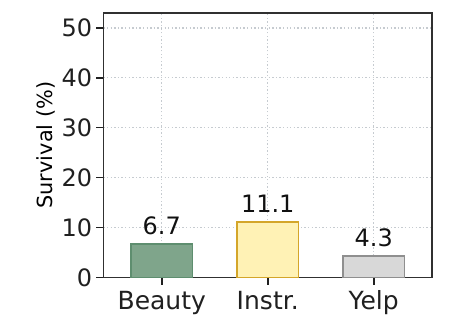}
\par\vspace{-1.5mm}
\makebox[\linewidth][l]{\hspace*{0.23\linewidth}\makebox[0.73\linewidth][c]{\scriptsize Top-10}}
\end{minipage}
\end{minipage}

\par\vspace{0.7mm}
\begin{minipage}[t]{0.515\linewidth}
\centering
{\scriptsize\textbf{(a)} Early Prefix Pruning}
\end{minipage}\hfill
\begin{minipage}[t]{0.470\linewidth}
\centering
{\scriptsize\textbf{(b)} Ground-Truth Survival}
\end{minipage}
\caption{Early beam pruning and ground-truth path survival.
(a) A target can survive at $c_1^*$ yet be irreversibly lost when $c_{2}^*$ is removed at the next beam step. Through joint-prefix alignment, PATH aims to retain relevant target paths that would otherwise be pruned, while allowing irrelevant branches to be discarded.
(b) Ground-truth prefix survival after $c_1$ and $c_2$, full-SID survival, and Hit@10 for the NTP baseline with $B=20$ on Beauty, Instruments, and Yelp. 
Even when the first-token prefix survives, many target paths are lost at the second decoding step. The subsequent decline from two-token prefix survival to full-SID survival is smaller in absolute terms.
This pattern motivates joint supervision of $(c_1,c_2)$ to preserve both early branches and their continuations. All four plots report percentages on the same vertical scale.}
\vspace{-5mm}
\label{fig:trie_pruning}
\end{figure}

Sequential recommendation aims to predict the next item a user will interact with from their interaction history~\citep{yang2023debiased,dang2026data,wang2025generative}. A generative approach to this task represents items as hierarchical semantic identifiers (SIDs) and predicts the next item by autoregressively generating its identifier~\citep{tiger,hou2025generating}.
To retrieve multiple candidate items, beam search extends identifier prefixes token by token, retaining only a limited set of high-scoring prefixes at each decoding step~\citep{oshima2026inference,wang2025proprag}.
Once a prefix is discarded, all items whose identifiers extend that prefix are excluded from further consideration (Figure~\ref{fig:trie_pruning}(a)). A target item can therefore enter the final candidate set only if its path survives both early branch selection and subsequent prefix expansion.

Recent work has improved SID construction and its integration with recommendation training~\citep{letter,lcrec,etegrec,diger}. Prefix retention has also become an explicit optimization target. APAO adaptively emphasizes vulnerable prefixes~\citep{apao26}, while BEAR encourages positive code tokens to remain within the Top-$B$ set~\citep{bear26}. These approaches primarily strengthen prefixes along the observed target path. 
We study how corpus-level transition statistics can complement this supervision by guiding predictions toward multiple plausible successor prefixes for the current history.

Our preliminary analysis reveals that improved first-token survival does not necessarily translate into better complete-path retention. Figure~\ref{fig:trie_pruning}(b) shows a sharp decline in ground-truth survival within the first two decoding steps at beam width $B=20$ across Beauty, Instruments, and Yelp. Targets pruned at these steps account for 91.9\%--96.6\% of all Hit@10 misses (Section~\ref{sec:survival}).
This concentration motivates supervision at early decoding steps, but the scope of that supervision matters.
In our first-token-only alignment diagnostic, first-token survival increases on all three datasets, while second-prefix and full-SID survival remain close to those of the next-token prediction baseline. These observations motivate supervision over early prefix combinations that couples first-token ($c_1$) branches with their second-token ($c_2$) continuations.

Historical transitions provide supervision over these early prefix combinations. Each next-token prediction (NTP) training instance directly supervises one observed successor, while transitions across the training corpus provide evidence about multiple plausible successors. 
We introduce Prefix Alignment with Temporal History (PATH) to incorporate this evidence into autoregressive training. PATH estimates transition distributions between two-token SID prefixes and aggregates the successor distributions associated with recent interactions using exponentially decaying weights. 
The resulting joint soft target over $(c_1,c_2)$ retains the association between each first-token branch and its second-token continuations. PATH aligns the model's joint prediction with this history-conditioned target through forward KL divergence, complementing NTP supervision along the observed target path. For efficient computation, we decompose the joint KL by the chain rule and estimate its conditional term with Monte Carlo sampling. This yields unbiased estimates of both the auxiliary objective and its gradient without evaluating second-token predictions for every first-token branch.

The same transition statistics also provide a supplementary ranking signal after candidate generation. Raw transition probabilities reflect both history-specific associations and global prefix frequency. We use pointwise mutual information (PMI) calibration to measure association relative to a prefix's overall prevalence and combine the resulting signal with the model score. This step reorders the completed beam without changing which candidates survive.

Our contributions are as follows:
\begin{itemize}\itemsep2pt
    \item We show that improving first-token survival alone yields little gain in full-SID survival across three datasets, motivating supervision of early prefix continuations.

    \item We propose PATH, which aligns early predictions with history-conditioned joint-prefix targets through unbiased stochastic KL estimation and supplements candidate ranking with PMI calibration.

    \item Through empirical evaluation, we demonstrate PATH's effectiveness in preserving target paths during decoding and improving recommendation performance across three datasets.
\end{itemize}

\section{Preliminaries}
\label{sec:prelim}

\subsection{Sequential Recommendation with Semantic Identifiers}

Given a user's interaction history
$\mathcal{H}=(v_1,\ldots,v_n)$ over an item set
$\mathcal{V}$, sequential recommendation predicts
the next item $v^*=v_{n+1}$.
A fixed tokenizer maps each item $v$ to a length-$L$
SID $c^v=(c_1^v,\ldots,c_L^v)$, where $L\geq2$ and
$c_l^v\in C_l$. We assume complete SIDs uniquely
identify items. Let $c_{1:l}^v$ denote a length-$l$
prefix and $c_{12}^v=(c_1^v,c_2^v)$ the joint prefix.

Let $X=(c^{v_1},\ldots,c^{v_n})$ and $c^*=c^{v^*}$.
The autoregressive model and its per-instance
next-token prediction (NTP) loss are
\begin{align}
p_\theta(c^*\mid X)
&=\prod_{l=1}^{L}
p_\theta(c_l^*\mid X,c_{<l}^*),
\label{eq:ar}\\
\mathcal{L}_{\mathrm{NTP}}
&=-\sum_{l=1}^{L}
\log p_\theta(c_l^*\mid X,c_{<l}^*),
\label{eq:ntp}
\end{align}
where $c_{<l}^*$ contains the preceding target tokens.

\subsection{Trie-Constrained Decoding and Prefix Survival}

Trie-constrained beam search expands prefixes through
valid trie edges and retains up to $B$ highest-scoring
extensions at each step. Let $\mathcal{B}_l(X)$ denote
the retained length-$l$ prefixes and
$\mathcal{B}(X)\subseteq\mathcal{V}$ the completed
item candidates. Once a target prefix is pruned,
subsequent expansions cannot recover the target.

For an evaluation set
$\mathcal{D}=\{(X_i,v_i^*)\}_{i=1}^{M}$,
the level-$l$ prefix survival rate is
\begin{equation}
S_{c_l}
=
\frac{1}{M}\sum_{i=1}^{M}
\mathbf{1}
\left[c_{1:l}^{v_i^*}\in\mathcal{B}_l(X_i)\right],
\label{eq:prefix_survival}
\end{equation}
where $\mathbf{1}[\cdot]$ is the indicator function.
Full-SID survival is $S_{\mathrm{full}}=S_{c_L}$,
and irreversible pruning implies
\begin{equation}
1\geq S_{c_1}\geq S_{c_2}\geq\cdots
\geq S_{c_L}=S_{\mathrm{full}}.
\label{eq:survival_monotonicity}
\end{equation}

Full-SID survival measures candidate inclusion, whereas Hit@$K$ additionally depends on ranking, with $\mathrm{Hit}@K\leq S_{\mathrm{full}}$. Reranking preserves the candidate set and all survival rates.

\section{PATH: Joint-Prefix Alignment and PMI Calibration}
\label{sec:method}

\begin{figure}[t]
\centering
\includegraphics[width=\linewidth]
{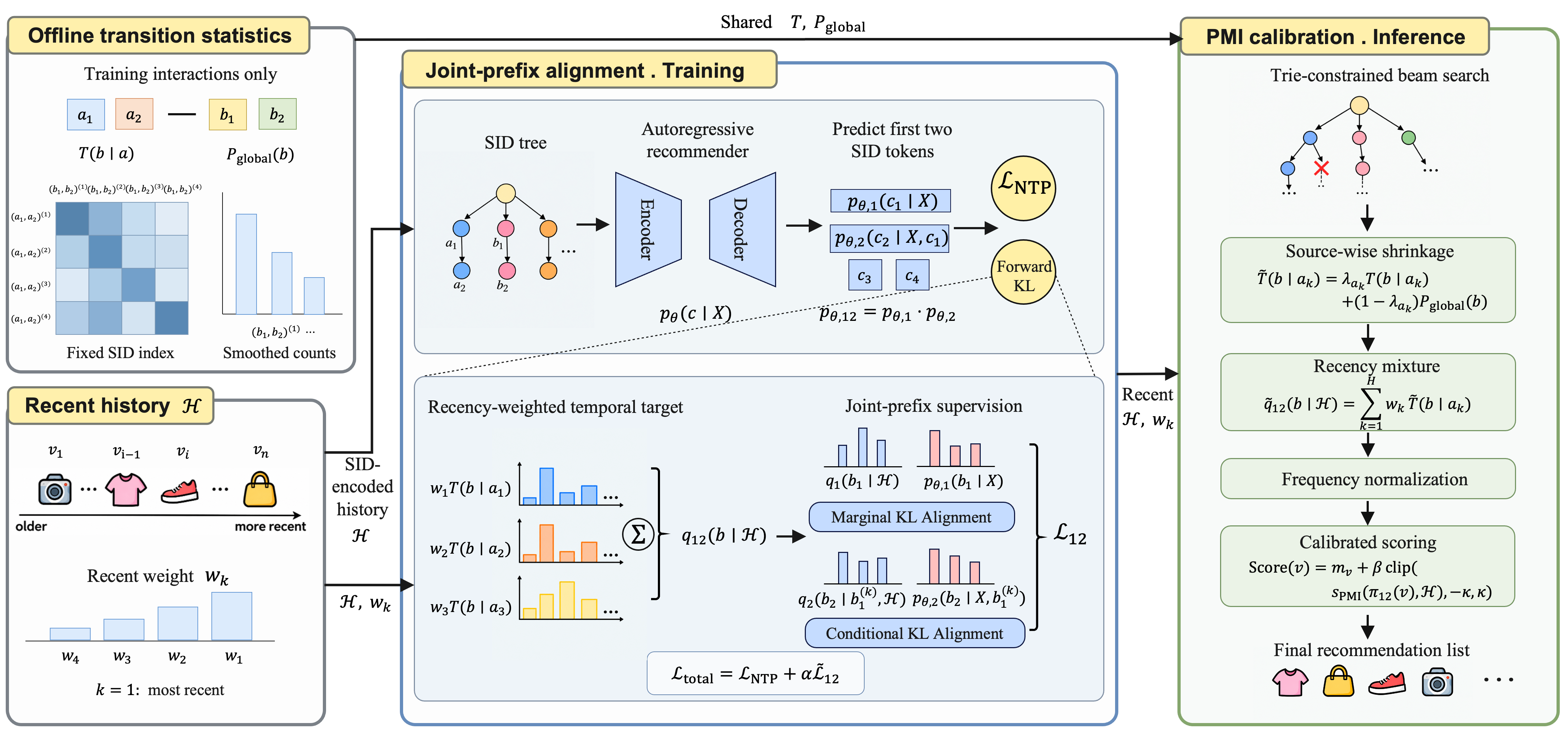}
\vspace{-5mm}
\caption{Overview of PATH.
Joint-prefix transitions estimated from training
interactions are aggregated over recent history
to construct a temporal soft target.
Joint forward-KL alignment supervises early decoding
paths, while PMI-based calibration adjusts completed
candidate scores relative to global prefix frequency.}
\vspace{-3mm}
\label{fig:framework}
\end{figure}

PATH combines history-conditioned joint-prefix alignment
during training with PMI-based candidate calibration
during inference (Figure~\ref{fig:framework}).

\subsection{Joint-Prefix Transition Statistics}
\label{sec:transition}

This section uses two-token prefixes to retain these
branch--continuation relationships while pooling
interactions among items sharing a prefix, targeting
the positions where our analysis finds most early
target pruning (section~\ref{sec:survival}).

Define $\pi_{12}(v)=(c_1^v,c_2^v)$, and let
$\mathcal{P}_{12}\subseteq C_1\times C_2$ denote the
valid two-token prefixes induced by the fixed SID index.
Transition counts are estimated exclusively from
training interactions.
We use $a=(a_1,a_2)$ and $b=(b_1,b_2)$ to denote
source and destination joint prefixes, respectively.
For each adjacent item pair $(v_t,v_{t+1})$, we count
a transition from $a=\pi_{12}(v_t)$ to
$b=\pi_{12}(v_{t+1})$.
Let $C(a,b)$ be its count across training sequences and
$C_{\mathrm{out}}(a)=
\sum_{b'\in\mathcal{P}_{12}}C(a,b')$.
The smoothed transition distribution is
\begin{equation}
T(b\mid a)
=
\frac{C(a,b)+\varepsilon}
{C_{\mathrm{out}}(a)+\varepsilon|\mathcal{P}_{12}|},
\label{eq:trans}
\end{equation}
where $\varepsilon>0$ provides additive smoothing.

We define the source-prefix distribution and the
corresponding reference marginal as
\begin{equation}
\begin{aligned}
P_{\mathrm{src}}(a)
=
\frac{C_{\mathrm{out}}(a)}
{\sum_{a'\in\mathcal{P}_{12}}C_{\mathrm{out}}(a')},
&&
P_{\mathrm{global}}(b)=
\sum_{a\in\mathcal{P}_{12}}
P_{\mathrm{src}}(a)T(b\mid a).
\end{aligned}
\label{eq:global_prior}
\end{equation}
Thus, $P_{\mathrm{global}}$ describes destination-prefix
frequency under the smoothed transition model.

\subsection{Recency-Weighted Temporal Targets}
\label{sec:history}

Conditioning only on the last interaction makes the
target sensitive to a single observation, while equal
weighting fails to distinguish recent interactions
from older ones. We aggregate a recent history window
with exponential decay to retain several sources of
transition evidence while emphasizing recency.

For a nonempty history $\mathcal{H}=(v_1,\ldots,v_n)$,
let $H=\min(n,K_{\mathrm{hist}})$.
We construct the temporal soft target over destination
joint prefixes by mixing the corresponding transition
distributions
\begin{align}
q_{12}(b\mid\mathcal{H})
=
\sum_{k=1}^{H}
w_k\,T\!\left(
b\mid\pi_{12}(v_{n-k+1})
\right),
\label{eq:softtarget}
\end{align}
where $w_k=\frac{\gamma^{k-1}}{\sum_{j=1}^{H}\gamma^{j-1}}$is the normalized weight of the $k$-th most recent item, $ 0<\gamma\leq1, k \in \{1,\ldots,H\}$.
For $\gamma<1$, recent interactions receive greater
weight, while earlier interactions within the window
retain a nonzero contribution.
When the history is empty, we set
$q_{12}(b\mid\mathcal{H})=P_{\mathrm{global}}(b)$.

\subsection{Joint Forward-KL Alignment}
\label{sec:training}

PATH aligns the joint two-token distribution to supervise both branch selection and within-branch continuations.

Given the SID-encoded history $X$, the model assigns
probability to a destination prefix $b=(b_1,b_2)$ as
$
p_{\theta,12}(b_1,b_2\mid X)
=
p_{\theta,1}(b_1\mid X)\,
p_{\theta,2}(b_2\mid X,b_1)
\label{eq:model_joint}
$, 
where $p_{\theta,1}$ and $p_{\theta,2}$ are the
first- and second-token predictive distributions.
Writing $q_{12}(b_1,b_2\mid\mathcal{H})$ for
$q_{12}((b_1,b_2)\mid\mathcal{H})$, we extend the
target by zero outside $\mathcal{P}_{12}$ and define
\begin{align}
q_1(b_1\mid\mathcal{H})
=
\sum_{b_2\in C_2}
q_{12}(b_1,b_2\mid\mathcal{H}), \quad
q_2(b_2\mid b_1,\mathcal{H})
=
\frac{q_{12}(b_1,b_2\mid\mathcal{H})}
{q_1(b_1\mid\mathcal{H})},
\label{eq:target_conditional}
\end{align}
with the conditional defined for
$q_1(b_1\mid\mathcal{H})>0$.
Below, we suppress conditioning on $\mathcal{H}$ and $X$.

By the chain rule, the joint forward KL decomposes as
\begin{equation}
\begin{aligned}
\mathcal{L}_{12}
&=
\mathrm{KL}\!\left(
q_{12}(\cdot\mid\mathcal{H})
\,\|\,p_{\theta,12}(\cdot\mid X)
\right)\\
&=
\mathrm{KL}\!\left(
q_1(\cdot\mid\mathcal{H})
\,\|\,p_{\theta,1}(\cdot\mid X)
\right)\\
&\quad+
\mathbb{E}_{b_1\sim q_1(\cdot\mid\mathcal{H})}
\left[
\mathrm{KL}\!\left(
q_2(\cdot\mid b_1,\mathcal{H})
\,\|\,p_{\theta,2}(\cdot\mid X,b_1)
\right)
\right].
\end{aligned}
\label{eq:chain_kl}
\end{equation}
The first term aligns probabilities across first-token
branches; the second aligns their continuations,
weighted by the target branch probabilities.
Forward KL penalizes insufficient model probability
on prefix combinations carrying substantial target
mass, encouraging coverage of empirically plausible
paths.

The marginal term is computed exactly in the standard
forward pass. Evaluating the conditional term exactly
would require second-token predictions for every
supported first-token branch.

We draw $K_{\mathrm{mc}}$ independent samples
$b_1^{(k)}\sim q_1$ with replacement and estimate
\begin{equation}
\begin{aligned}
\widetilde{\mathcal{L}}_{12}
&=\mathrm{KL}(q_1\|p_{\theta,1})\\
&\quad+\frac{1}{K_{\mathrm{mc}}}
\sum_{k=1}^{K_{\mathrm{mc}}}
\mathrm{KL}\!\left(
q_2(\cdot\mid b_1^{(k)})
\|p_{\theta,2}(\cdot\mid b_1^{(k)})
\right).
\end{aligned}
\label{eq:mc_kl}
\end{equation}
Sampling from $q_1$ yields an unbiased estimate of
$\mathcal{L}_{12}$. Since the target is independent
of $\theta$, the gradient estimate is also unbiased
(Appendix~\ref{sec:mc_proof}).

The complete training objective is
\begin{equation}
\mathcal{L}_{\mathrm{total}}
=
\mathcal{L}_{\mathrm{NTP}}
+
\alpha\,\widetilde{\mathcal{L}}_{12},
\label{eq:total}
\end{equation}
where $\alpha\geq0$ controls alignment strength.
The auxiliary loss supervises the first two decoding
positions, while NTP remains active over the complete
observed SID.
The tokenizer and recommender architecture are unchanged.

\subsection{PMI-Based Candidate Calibration}
\label{sec:inference}

This section calibrates candidate scores using history-specific transition evidence relative to global prefix frequency. To reduce sensitivity to sparse counts, we shrink each source-conditioned distribution toward $P_{\mathrm{global}}$. Let $a_k=\pi_{12}(v_{n-k+1})$; we define
\begin{equation}
\begin{aligned}
\widetilde{T}(b\mid a_k)
&=
\lambda_{a_k}T(b\mid a_k)
+
(1-\lambda_{a_k})P_{\mathrm{global}}(b),
\qquad
\widetilde{q}_{12}(b\mid\mathcal{H})
&=
\sum_{k=1}^{H}w_k\widetilde{T}(b\mid a_k).
\end{aligned}
\label{eq:shrinkage}
\end{equation}
Here,
$\lambda_{a_k}
=
C_{\mathrm{out}}(a_k)/
\bigl(C_{\mathrm{out}}(a_k)+\tau\bigr)$
is a support-dependent reliability weight, and $\tau>0$ controls
the shrinkage strength. A frequently observed source prefix receives
more weight from its transition distribution, whereas a prefix with
limited support relies more on the global marginal. When
$C_{\mathrm{out}}(a_k)=0$, its adjusted distribution reduces to
$P_{\mathrm{global}}$. During training, such a source contributes the uniform
transition row defined by Equation~\eqref{eq:trans}.
For an empty history, we set
$\widetilde{q}_{12}(\cdot\mid\mathcal{H})
=P_{\mathrm{global}}$, yielding zero calibration.

The calibration signal is
\begin{equation}
s_{\mathrm{PMI}}(b,\mathcal{H})
=
\log
\frac{\widetilde{q}_{12}(b\mid\mathcal{H})}
{P_{\mathrm{global}}(b)}.
\label{eq:pmi}
\end{equation}
The signal is positive when the destination prefix is more likely
under the current history than under its global reference frequency.
It becomes zero when all source prefixes in the recent history have
no observed outgoing transitions.

For each completed candidate $v\in\mathcal{B}(X)$,
we use its sequence log-probability
$m_v=\log p_\theta(c^v\mid X)$ as the base score
for calibration
\begin{equation}
\mathrm{Score}(v)
=
m_v
+
\beta\,
\operatorname{clip}
\left(
s_{\mathrm{PMI}}\!\left(\pi_{12}(v),\mathcal{H}\right),
-\kappa,\kappa
\right),
\label{eq:score}
\end{equation}
where $\beta\geq0$ controls the calibration strength and
$\kappa>0$ bounds the magnitude of the PMI adjustment.
Calibration only reranks the completed candidates and therefore
leaves the candidate set and survival rates unchanged. Candidates
sharing the same joint prefix receive the same adjustment.

\paragraph{Computational cost.}
Let $P=|\mathcal{P}_{12}|$.
An explicit dense transition table requires $O(P^2)$
storage, and directly constructing a history target
costs $O(HP)$.
Monte Carlo estimation reduces conditional model
evaluations to $K_{\mathrm{mc}}$ sampled branches.
Calibration requires $O(BH)$ table lookups and
$O(B\log B)$ sorting, with no additional
autoregressive decoding.

\section{Experiments}
\label{sec:exp}

We evaluate PATH from three perspectives: overall effectiveness, component contributions, and consistency with the proposed pruning mechanism. Main experiments compare recommendation performance across three benchmarks. Ablations isolate joint-prefix training and PMI calibration. Survival diagnostics, structure controls, and beam-width sweeps test whether the observed gains follow the early-pruning hypothesis, while sensitivity curves characterize the range of effective training and inference weights.

\subsection{Experimental Setup}
\label{sec:setup}

We evaluate on Amazon Beauty, Amazon Instruments, and
Yelp using leave-one-out splitting and full-catalog
Hit@10 and NDCG@10. PATH uses the T5-based
LETTER+TIGER framework~\citep{letter,tiger}, which also
provides our NTP baseline. We rerun LETTER+TIGER,
APAO~\citep{apao26}, and BEAR~\citep{bear26} using
their original settings, and the remaining baselines
using the settings reported in DIGER~\citep{diger}.
Our BEAR rerun fine-tunes Llama-3.2-3B.
Unless otherwise stated, PATH and its NTP baseline
use beam width $B=20$. All experiments use seed 42.
Calibration variants share the same checkpoint and
candidate set within each dataset. PATH's alignment
weight is selected on the validation set.
Dataset statistics, detailed evaluation protocols,
and implementation settings are provided in
Appendix~\ref{sec:appendix}.

\begin{table}[t]

\caption{Recommendation performance on three benchmarks.
H@$K$ and N@$K$ denote Hit@$K$ and NDCG@$K$, respectively.
Bold indicates the best result in each column.}
\label{tab:main}
\centering
\scriptsize
\setlength{\tabcolsep}{1.8pt}
\begin{tabular}{llcccccc}
\toprule
& & \multicolumn{2}{c}{Beauty}
  & \multicolumn{2}{c}{Instruments}
  & \multicolumn{2}{c}{Yelp} \\
\cmidrule(lr){3-4}
\cmidrule(lr){5-6}
\cmidrule(lr){7-8}
Category & Method
& H@10 & N@10
& H@10 & N@10
& H@10 & N@10 \\
\midrule

\multirow{6}{*}{\shortstack[l]{Sequential/\\collaborative}}
& MF~\citep{mf}
& .0474 & .0191 & .0735 & .0412 & .0381 & .0190 \\
& Caser~\citep{caser}
& .0347 & .0176 & .0710 & .0409 & .0263 & .0134 \\
& HGN~\citep{hgn}
& .0512 & .0266 & .1048 & .0774 & .0326 & .0159 \\
& LightGCN~\citep{lightgcn}
& .0511 & .0260 & .1000 & .0728 & .0407 & .0207 \\
& SASRec~\citep{sasrec}
& .0588 & .0313 & .0947 & .0690 & .0296 & .0152 \\
& BERT4Rec~\citep{bert4rec}
& .0347 & .0170 & .0822 & .0608 & .0291 & .0159 \\
\midrule

\multirow{4}{*}{Generative baselines}
& BIGRec~\citep{bigrec}
& .0299 & .0198 & .0576 & .0491 & .0169 & .0142 \\
& P5-SID~\citep{p5,diger}
& .0584 & .0335 & .0964 & .0730 & .0324 & .0170 \\
& P5-CID~\citep{p5,diger}
& .0597 & .0347 & .0987 & .0751 & .0347 & .0181 \\
& TIGER~\citep{tiger}
& .0610 & .0331 & .1058 & .0797 & .0407 & .0213 \\
\midrule

\multirow{6}{*}{\shortstack[l]{Tokenizer/decoding\\improvements}}
& STE~\citep{ste}
& .0134 & .0067 & .0554 & .0360 & .0147 & .0077 \\
& ETEGRec~\citep{etegrec}
& .0615 & .0335 & .1106 & .0810 & .0415 & .0214 \\
& LETTER~\citep{letter}
& .0671 & .0347 & .1107 & .0825 & .0429 & .0228 \\
& DIGER~\citep{diger}
& .0683 & .0372 & .1138 & .0844 & .0432 & .0227 \\
& APAO~\citep{apao26}
& .0641 & .0337 & .1092 & .0811 & .0409 & .0216 \\
& BEAR~\citep{bear26}
& .0817 & .0444 & .0988 & .0737 & \textbf{.0577} & \textbf{.0332} \\
\midrule

Ours & PATH (no reranking)
& .0844 & .0477 & .1167 & .0874 & .0515 & .0286 \\
Ours & PATH + PMI calibration
& \textbf{.0856} & \textbf{.0484}
& \textbf{.1176} & \textbf{.0881}
& .0525 & .0293 \\
\bottomrule
\end{tabular}
\vspace{-3mm}
\end{table}

\subsection{Main Results}
\label{sec:main}

Table~\ref{tab:main} reports recommendation performance
across the three benchmarks.
PATH with PMI calibration achieves the best Hit@10
and NDCG@10 on Beauty and Instruments, while BEAR
performs best on Yelp.
PATH reaches NDCG@10 values of 0.0484, 0.0881,
and 0.0293, improving over LETTER+TIGER by
39.5\%, 6.7\%, and 28.6\%, respectively.
Compared with BEAR, PATH improves NDCG@10 by
approximately 9.0\% on Beauty and 19.5\% on Instruments,
with corresponding Hit@10 gains of 4.8\% and 19.0\%.

PATH outperforms BEAR on Beauty and Instruments but trails it on Yelp. This comparison involves different backbones and pretraining configurations: BEAR fine-tunes Llama-3.2-3B, while PATH uses the T5-based LETTER+TIGER framework. BEAR's model capacity and pretraining may contribute to more effective modeling of Yelp interaction histories and partly explain the observed gap. PATH also achieves substantially faster inference than the BEAR configuration evaluated here, while its PMI calibration requires no additional autoregressive decoding. These comparisons reflect differences in backbone, pretraining, and training objectives.

Joint-prefix training accounts for most of PATH's
improvement. Without calibration, PATH achieves
NDCG@10 values of 0.0477, 0.0874, and 0.0286
on Beauty, Instruments, and Yelp, respectively.
PMI calibration further improves ranking within
the same candidate sets. Compared with DIGER, calibrated PATH improves NDCG@10 by
30.2\%, 4.4\%, and 28.9\%, respectively. It also
achieves higher Hit@10 and NDCG@10 than our APAO
rerun on all three datasets. We further examine
path retention and component contributions in
sections~\ref{sec:survival} and~\ref{sec:ablation},
respectively.

\subsection{Early Pruning and Joint-Prefix Survival}
\label{sec:survival}

We first examine where targets leave the beam and
whether improving the first decoding step translates
into complete-path retention.
For the NTP baseline with $B=20$, first-token survival
is 43.57\%, 47.02\%, and 28.87\% on Beauty,
Instruments, and Yelp, respectively.
By the second step, these rates fall to
14.26\%, 16.83\%, and 7.57\%.
The fraction of Hit@10 misses whose targets are pruned within the first two decoding steps is
\begin{equation}
\rho_{\mathrm{early}}
=
\frac{1-S_{c_2}}{1-\mathrm{Hit}@10},
\label{eq:early_failure_share}
\end{equation}
which yields 91.9\%, 93.5\%, and 96.6\% on the
three datasets.
These results identify the first two positions
as the main location of target-path loss.

Table~\ref{tab:prefix_survival} compares NTP,
$c_1$-only alignment, and joint alignment.
The $c_1$-only variant improves first-token survival
on all three datasets, while second-prefix and
full-SID survival remain close to the baseline.
On Beauty, for example, $S_{c_1}$ increases from
43.57\% to 46.39\%, but $S_{\mathrm{full}}$
changes only from 10.07\% to 10.10\%.

Joint alignment achieves lower $S_{c_1}$ than
$c_1$-only alignment but improves both subsequent
stages on every dataset.
Its full-SID survival reaches 11.75\%, 14.37\%,
and 7.67\%, compared with 10.07\%, 13.84\%,
and 6.69\% for NTP.
Thus, the configuration with the highest
first-token survival does not produce the best
complete-path retention.
This supports supervising second-token continuations
alongside first-token branches, as motivated in
section~\ref{sec:intro}.
Complete survival and ranking diagnostics
are provided in Table~\ref{tab:survival} in
Appendix~\ref{sec:additional_survival}.

\begin{table}[t]
\caption{Prefix and full-SID survival (\%) at $B=20$. Bold indicates the best result within each dataset.}
\label{tab:prefix_survival}
\centering
\scriptsize
\setlength{\tabcolsep}{3pt}
\begin{tabular}{llcccc}
\toprule
Dataset & Training objective & $\alpha$
& $S_{c_1}$ & $S_{c_2}$ & $S_{\mathrm{full}}$ \\
\midrule

\multirow{3}{*}{Beauty}
& NTP & 0
& 43.57 & 14.26 & 10.07 \\
& $c_1$-only alignment & 0.4
& \textbf{46.39} & 14.13 & 10.10 \\
& Joint alignment & 0.2
& 45.24 & \textbf{15.76} & \textbf{11.75} \\
\midrule

\multirow{3}{*}{Instruments}
& NTP & 0
& 47.02 & 16.83 & 13.84 \\
& $c_1$-only alignment & 0.2
& \textbf{50.77} & 16.79 & 13.80 \\
& Joint alignment & 0.2
& 49.27 & \textbf{17.75} & \textbf{14.37} \\
\midrule

\multirow{3}{*}{Yelp}
& NTP & 0
& 28.87 & 7.57 & 6.69 \\
& $c_1$-only alignment & 0.3
& \textbf{32.60} & 7.55 & 6.69 \\
& Joint alignment & 0.3
& 31.06 & \textbf{8.58} & \textbf{7.67} \\
\bottomrule
\end{tabular}
\vspace{-3mm}
\end{table}

\subsection{History-Conditioned Supervision Controls}
\label{sec:control}

We assess the contribution of history matching on Beauty using the same joint-KL objective, training recipe, $\alpha=0.2$, and $K_{\mathrm{mc}}=2$, with calibration disabled. 
The global-marginal control replaces the temporal target with $P_{\mathrm{global}}$.
The history-shuffled control cyclically shifts targets by one position within each batch, preserving the target collection while disrupting its correspondence with input histories.

As shown in Table~\ref{tab:control}, both controls improve NDCG@10 over NTP but reduce full-SID survival, whereas PATH improves both. At smaller alignment weights, the controls can also improve survival, but PATH achieves higher NDCG@10 across the reported settings (Table~\ref{tab:control_sweep} in Appendix~\ref{sec:additional_survival}).

\begin{table}[t]
\caption{History-conditioned supervision controls on Beauty. Auxiliary losses use $\alpha=0.2$, with inference calibration disabled. Survival rates are percentages. Bold indicates the best result in each column.}
\label{tab:control}
\centering
\scriptsize
\setlength{\tabcolsep}{3pt}
\begin{tabular}{lccccc}
\toprule
Training target
& $S_{c_1}$ & $S_{c_2}$ & $S_{\mathrm{full}}$
& H@10 & N@10 \\
\midrule

None (NTP)
& 43.57 & 14.26 & 10.07 & .0671 & .0347 \\

Global marginal
& 40.95 & 13.56 & 9.44 & .0667 & .0368 \\

History shuffled
& 40.88 & 13.37 & 9.39 & .0669 & .0360 \\

History aligned (PATH)
& \textbf{45.24}
& \textbf{15.76}
& \textbf{11.75}
& \textbf{.0844}
& \textbf{.0477} \\

\bottomrule
\end{tabular}
\end{table}

\subsection{Training and Inference Ablations}
\label{sec:ablation}

We examine the contributions of joint training and
post-generation calibration.
Table~\ref{tab:ablation} compares NTP with PATH under
three inference settings: no calibration, a conditional
prior using $\log q_{12}$, and PMI calibration using
$\log(\widetilde{q}_{12}/P_{\mathrm{global}})$.
Within each dataset, all PATH variants share the
same checkpoint and completed candidate set.

\textbf{Joint training provides most of the gain.}
Without calibration, PATH improves NDCG@10 over NTP
by 37.5\%, 5.9\%, and 25.4\% on Beauty,
Instruments, and Yelp.
The survival analysis in Section~\ref{sec:survival}
also shows improved target-path retention.

\textbf{Calibration provides an additional ranking gain.}
PMI increases both Hit@10 and NDCG@10 over uncalibrated
PATH on every dataset.
The complete PMI calibration scheme also yields
slightly higher NDCG@10 than the conditional prior:
.0484 versus .0482 on Beauty,
.0881 versus .0875 on Instruments, and
.0293 versus .0292 on Yelp.
The advantage is metric-dependent: the conditional
prior achieves higher Hit@1 on Yelp.

Calibration improves ranking within a fixed candidate
set, leaving survival rates unchanged.
The average rank of surviving targets also improves
on all three datasets (Table~\ref{tab:survival} in
Appendix~\ref{sec:additional_survival}).

\begin{table}[t]
\caption{Training and calibration ablations. PATH calibration variants share the same checkpoint and candidate set within each dataset. Bold indicates the best result in each column.}
\label{tab:ablation}
\centering
\scriptsize
\setlength{\tabcolsep}{2.5pt}
\begin{tabular}{llccc|ccc|ccc}
\toprule
& & \multicolumn{3}{c|}{Beauty}
  & \multicolumn{3}{c|}{Instruments}
  & \multicolumn{3}{c}{Yelp} \\
\cmidrule(lr){3-5}
\cmidrule(lr){6-8}
\cmidrule(lr){9-11}
Training & Calibration
& H@1 & H@10 & N@10
& H@1 & H@10 & N@10
& H@1 & H@10 & N@10 \\
\midrule
NTP & None
& .0118 & .0671 & .0347
& .0623 & .1107 & .0825
& .0081 & .0429 & .0228 \\
\midrule
\multirow{3}{*}{PATH}
& None
& .0204 & .0844 & .0477
& .0658 & .1167 & .0874
& .0117 & .0515 & .0286 \\
& Conditional prior
& .0206 & .0850 & .0482
& \textbf{.0664} & .1163 & .0875
& \textbf{.0123} & .0524 & .0292 \\
& PMI
& \textbf{.0208} & \textbf{.0856} & \textbf{.0484}
& \textbf{.0664} & \textbf{.1176} & \textbf{.0881}
& .0118 & \textbf{.0525} & \textbf{.0293} \\
\bottomrule
\end{tabular}
\end{table}

\begin{figure}[t]
\centering
\includegraphics[width=0.75\linewidth]
{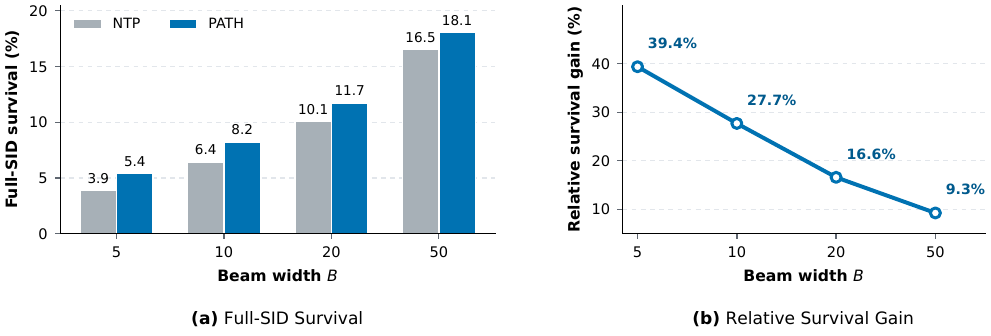}
\vspace{-4mm}
\caption{Beam-width analysis on Beauty using fixed
NTP and PATH checkpoints.
(a) Full-SID survival at $B\in\{5,10,20,50\}$.
(b) Relative survival gain of PATH over NTP.
PATH improves survival at every tested width,
with the largest relative gain at $B=5$.}
\label{fig:beamsweep}
\vspace{-5mm}
\end{figure}

\subsection{Beam-Width Analysis}
\label{sec:beamsweep}

We fix the NTP and PATH checkpoints on Beauty and vary $B\in\{5,10,20,50\}$. PATH improves full-SID survival at every tested width (Figure~\ref{fig:beamsweep}).
Relative gains range from 39.4\% at $B=5$ to 9.3\% at $B=50$, while absolute gains remain between 1.52 and 1.79 percentage points without a monotonic trend. Thus, narrower beams yield larger proportional improvements. Complete results appear in Appendix~\ref{sec:additional_survival}, Table~\ref{tab:beamsweep}.

\section{Related Work}
\label{sec:related}

\textbf{Generative sequential recommendation.}
Sequential recommenders such as GRU4Rec~\citep{grurec},
SASRec~\citep{sasrec}, and BERT4Rec~\citep{bert4rec}
learn to predict items from interaction histories.
P5~\citep{p5} and M6-Rec~\citep{m6rec} formulate
recommendation tasks through language-based frameworks.
TIGER~\citep{tiger} combines RQ-VAE-based semantic
identifiers with autoregressive generation, providing
the SID-based retrieval framework used in this work.

\textbf{Semantic identifiers and recommendation alignment.}
LETTER~\citep{letter} incorporates semantic,
collaborative, and code-diversity regularization
into item tokenization.
LC-Rec~\citep{lcrec} combines vector-quantized item
indices with alignment tasks to integrate language
and collaborative semantics.
ETEGRec~\citep{etegrec} couples tokenization and
recommendation through alignment objectives and
alternating optimization, while DIGER~\citep{diger}
enables recommendation gradients to guide SID learning
through differentiable tokenization.
These methods improve item representations or their
coordination with recommendation learning.
PATH uses fixed SIDs and adds joint-prefix supervision
and candidate calibration to the LETTER+TIGER framework.

\textbf{Constrained decoding and prefix-aware optimization.}
Generative retrieval produces identifiers or retrieval
keys autoregressively~\citep{dsi,seal,genrie}.
Trie constraints restrict generation to valid
identifiers, and STATIC~\citep{static26} accelerates
this process through vectorized trie traversal.
To address premature pruning, APAO~\citep{apao26}
adaptively emphasizes vulnerable target prefixes,
while BEAR~\citep{bear26} regularizes positive-token
rankings using a relaxed necessary condition for
beam retention.
These methods focus on retaining the observed target
path. PATH complements target-path supervision with
a history-conditioned distribution over multiple
joint prefixes, supervising both first-token branches
and their second-token continuations.

\textbf{Collaborative transition signals.}
Item associations and sequential transitions are
established recommendation signals, as used in
item-based collaborative filtering~\citep{itemknn}
and factorized personalized Markov chains~\citep{fpmc}.
Recent work also supplies collaborative cues to SID
generation through personalized natural language
at inference~\citep{collabinj26}.
PATH aggregates training-corpus transitions at the
joint-prefix level to construct history-conditioned
soft targets. It reuses these statistics for
post-generation calibration relative to global prefix
frequency, connecting collaborative transition
modeling with both autoregressive training and
candidate ranking.

\section{Conclusion and Limitations}
\label{sec:conclusion}

We identify a gap between first-token survival
and complete-path retention in generative
recommendation.
PATH addresses this gap by aligning early
autoregressive predictions with history-conditioned
joint-prefix targets derived from training
transitions.
An unbiased stochastic estimator makes this
alignment tractable, while PMI calibration reuses
the transition evidence to refine candidate ranking.
Experiments across three benchmarks show improved
path retention and recommendation accuracy, with
most gains coming from joint-prefix training.
Controls on Beauty further support the value of
matching transition supervision to the current
history.
These findings motivate supervising early branches
together with their continuations when learning
to retrieve items through autoregressive decoding.

\textbf{Limitations.} First, the transition model uses first-order contextual transitions; higher-order and multihop temporal behavior remain open directions. Second, the experiments cover three medium-scale public datasets, and generalization to very large industrial catalogs requires further study.

\bibliography{references}
\bibliographystyle{iclr2027_conference}

\clearpage
\tableofcontents

\clearpage
\appendix

\section{Datasets and Implementation Details}
\label{sec:appendix}

\begin{table}[h]
\caption{Dataset statistics before validation and
test holdout. Average length denotes the mean
number of interactions per user.}
\label{tab:dataset_stats}
\centering
\scriptsize
\setlength{\tabcolsep}{3pt}
\begin{tabular}{lrrrrr}
\toprule
Dataset & Users & Items & Interactions
& Avg. length & Density \\
\midrule

Beauty
& 22{,}363 & 12{,}101 & 198{,}502
& 8.88 & 0.0734\% \\

Instruments
& 24{,}772 & 9{,}922 & 206{,}153
& 8.32 & 0.0839\% \\

Yelp
& 30{,}431 & 20{,}033 & 316{,}354
& 10.40 & 0.0519\% \\

\bottomrule
\end{tabular}
\end{table}

\paragraph{Dataset statistics.}
Table~\ref{tab:dataset_stats} summarizes the datasets
before validation and test holdout.
Average sequence length is the number of interactions
divided by the number of users.
Density is the number of interactions divided by
the product of the numbers of users and items.

\paragraph{Evaluation protocol.}
For each user, the penultimate interaction is
reserved for validation and the final interaction
for testing.
Transition statistics are computed exclusively
from training interactions.
We report full-catalog Hit@10 and NDCG@10;
with one held-out target per user, Hit@10 equals
Recall@10.
PATH and its NTP baseline use trie-constrained
beam search with $B=20$, unless otherwise stated.
The beam-width analysis evaluates
$B\in\{5,10,20,50\}$.

\paragraph{Implementation and hyperparameters.}
LETTER+TIGER combines RQ-VAE semantic identifiers
with a T5 recommender and uses codebook size $K=256$.
The tokenizer remains fixed during recommender
training.
The NTP baseline disables both alignment and
calibration ($\alpha=0,\beta=0$).

Table~\ref{tab:implementation} lists the main PATH settings.
We select $\alpha$ separately for each dataset on
the validation set from
$\{0.1,0.2,0.3,0.4,0.5\}$.
The history window and decay factor are shared
by training-target construction and inference
calibration.
Models are trained independently across datasets,
and all experiments conducted in this work use
seed 42.
Supplementary experiments specify any departures
from the main configuration alongside their results.

\begin{table}[h]
\caption{PATH hyperparameters for the main experiments.
Dataset-specific alignment weights are listed in
the order Beauty, Instruments, and Yelp.}
\label{tab:implementation}
\centering
\small
\setlength{\tabcolsep}{8pt}
\begin{tabular}{ll}
\toprule
Parameter & Main setting \\
\midrule
History window $K_{\mathrm{hist}}$ & 5 \\
History decay $\gamma$ & 0.7 \\
Transition smoothing $\varepsilon$ & $10^{-4}$ \\
Monte Carlo samples $K_{\mathrm{mc}}$ & 2 \\
Alignment weight $\alpha$ & 0.2 / 0.2 / 0.3 \\
Calibration weight $\beta$ & 0.02 \\
Shrinkage strength $\tau$ & 30 \\
PMI clipping threshold $\kappa$ & 3.0 \\
\bottomrule
\end{tabular}
\end{table}

\section{Additional Survival and Control Results}
\label{sec:additional_survival}

\paragraph{Complete survival and ranking diagnostics.}
Table~\ref{tab:survival} extends the main survival comparison
with recommendation metrics and the average target rank
conditioned on full-SID survival.
The joint-alignment and PMI rows use the same checkpoint
within each dataset, so their survival rates coincide.
Calibration affects the ranks of retained targets
and the final recommendation metrics.

\begin{table}[h]
\caption{Complete survival and ranking diagnostics at
$B=20$. Survival rates are percentages.
Rank denotes the average target rank conditioned on
full-SID survival, for which lower is better.
The PMI row calibrates the joint-alignment checkpoint
immediately above it.
Bold indicates the best value within each dataset.}
\label{tab:survival}
\centering
\scriptsize
\setlength{\tabcolsep}{3pt}
\begin{tabular}{llcccccc}
\toprule
Dataset & Configuration
& $S_{c_1}$ & $S_{c_2}$ & $S_{\mathrm{full}}$
& Rank & Hit@10 & NDCG@10 \\
\midrule
\multirow{4}{*}{Beauty}
& NTP ($\alpha=0$)
& 43.57 & 14.26 & 10.07 & 8.13 & .0671 & .0347 \\
& $c_1$-only ($\alpha=0.4$)
& \textbf{46.39} & 14.13 & 10.10 & 7.81 & .0690 & .0377 \\
& Joint alignment ($\alpha=0.2$)
& 45.24 & \textbf{15.76} & \textbf{11.75}
& 7.21 & .0844 & .0477 \\
& Joint alignment + PMI
& 45.24 & \textbf{15.76} & \textbf{11.75}
& \textbf{7.11} & \textbf{.0856} & \textbf{.0484} \\
\midrule
\multirow{4}{*}{Instruments}
& NTP ($\alpha=0$)
& 47.02 & 16.83 & 13.84 & 5.31 & .1107 & .0825 \\
& $c_1$-only ($\alpha=0.2$)
& \textbf{50.77} & 16.79 & 13.80 & 5.40 & .1101 & .0818 \\
& Joint alignment ($\alpha=0.2$)
& 49.27 & \textbf{17.75} & \textbf{14.37}
& 5.13 & .1167 & .0874 \\
& Joint alignment + PMI
& 49.27 & \textbf{17.75} & \textbf{14.37}
& \textbf{5.02} & \textbf{.1176} & \textbf{.0881} \\
\midrule
\multirow{4}{*}{Yelp}
& NTP ($\alpha=0$)
& 28.87 & 7.57 & 6.69 & 8.29 & .0429 & .0228 \\
& $c_1$-only ($\alpha=0.3$)
& \textbf{32.60} & 7.55 & 6.69 & 8.23 & .0435 & .0234 \\
& Joint alignment ($\alpha=0.3$)
& 31.06 & \textbf{8.58} & \textbf{7.67}
& 7.82 & .0515 & .0286 \\
& Joint alignment + PMI
& 31.06 & \textbf{8.58} & \textbf{7.67}
& \textbf{7.61} & \textbf{.0525} & \textbf{.0293} \\
\bottomrule
\end{tabular}
\end{table}

\paragraph{Control-weight sweeps.}
Table~\ref{tab:control_sweep} reports the complete
alignment-weight sweeps for the global-marginal
and history-shuffled controls on Beauty.
All auxiliary objectives use the same forward-KL
formulation and $K_{\mathrm{mc}}=2$.
Both controls provide gains at low weights but
degrade at $\alpha=0.4$.
History-aligned PATH achieves higher NDCG@10 across
its reported settings.
The calibrated row connects these diagnostics to
the main result; comparisons between target
distributions use the uncalibrated rows.

\begin{table}[h]
\caption{Control-weight sweeps on Beauty with
$B=20$ and seed 42.
$\Delta$N@10 denotes the relative NDCG@10 change
from the NTP baseline.
All rows except the final row disable inference
calibration. Survival rates are percentages.
Bold indicates the best result in each metric column.}
\label{tab:control_sweep}
\centering
\scriptsize
\setlength{\tabcolsep}{3pt}
\begin{tabular}{lccccccc}
\toprule
Target / configuration & $\alpha$
& $S_{c_1}$ & $S_{c_2}$ & $S_{\mathrm{full}}$
& Hit@10 & NDCG@10 & $\Delta$N@10 \\
\midrule
NTP
& 0 & 43.57 & 14.26 & 10.07 & .0671 & .0347 & 0.0\% \\
\midrule
\multirow{3}{*}{Global marginal}
& 0.1 & 42.80 & 14.30 & 10.22 & .0715 & .0382 & +10.2\% \\
& 0.2 & 40.95 & 13.56 & 9.44 & .0667 & .0368 & +6.1\% \\
& 0.4 & 37.28 & 11.75 & 7.63 & .0567 & .0296 & -14.7\% \\
\midrule
\multirow{3}{*}{History shuffled}
& 0.1 & 42.62 & 14.40 & 10.20 & .0724 & .0386 & +11.3\% \\
& 0.2 & 40.88 & 13.37 & 9.39 & .0669 & .0360 & +3.8\% \\
& 0.4 & 35.87 & 11.06 & 7.13 & .0512 & .0269 & -22.5\% \\
\midrule
\multirow{2}{*}{History aligned (PATH)}
& 0.2 & 45.24 & \textbf{15.76} & \textbf{11.75}
& .0844 & .0477 & +37.5\% \\
& 0.4 & \textbf{45.58} & 15.33 & 11.54
& .0806 & .0455 & +31.2\% \\
PATH + PMI
& 0.2 & 45.24 & \textbf{15.76} & \textbf{11.75}
& \textbf{.0856} & \textbf{.0484} & \textbf{+39.5\%} \\
\bottomrule
\end{tabular}
\end{table}

\paragraph{Complete beam-width results.}
Table~\ref{tab:beamsweep} provides the numerical results
underlying Figure~\ref{fig:beamsweep}.
Both models use fixed checkpoints on Beauty,
with $\alpha=0.2$ for PATH.
Wider beams increase survival for both models,
while PATH retains more targets at every tested width.

\begin{table}[h]
\caption{Prefix survival at different beam widths
on Beauty. Survival rates are percentages.
Relative gain refers to PATH's full-SID survival
relative to NTP at the same beam width.
Bold indicates higher survival within each pair.}
\label{tab:beamsweep}
\centering
\scriptsize
\setlength{\tabcolsep}{3pt}
\begin{tabular}{clcccc}
\toprule
Beam $B$ & Model
& $S_{c_1}$ & $S_{c_2}$ & $S_{\mathrm{full}}$
& Relative gain \\
\midrule

\multirow{2}{*}{5}
& NTP
& 27.38 & 7.12 & 3.86 & -- \\
& PATH
& \textbf{28.95}
& \textbf{8.49}
& \textbf{5.38}
& +39.4\% \\
\midrule

\multirow{2}{*}{10}
& NTP
& 35.02 & 10.19 & 6.44 & -- \\
& PATH
& \textbf{36.48}
& \textbf{11.76}
& \textbf{8.23}
& +27.7\% \\
\midrule

\multirow{2}{*}{20}
& NTP
& 43.57 & 14.26 & 10.07 & -- \\
& PATH
& \textbf{45.24}
& \textbf{15.76}
& \textbf{11.75}
& +16.6\% \\
\midrule

\multirow{2}{*}{50}
& NTP
& 55.84 & 21.40 & 16.54 & -- \\
& PATH
& \textbf{58.69}
& \textbf{22.84}
& \textbf{18.07}
& +9.3\% \\
\bottomrule
\end{tabular}
\end{table}

\section{Hyperparameter Sensitivity}
\label{sec:sensitivity}

\subsection{Alignment and Calibration Weights}
\label{sec:weight_sensitivity}

\begin{figure}[t]
\centering
\includegraphics[width=\linewidth]
{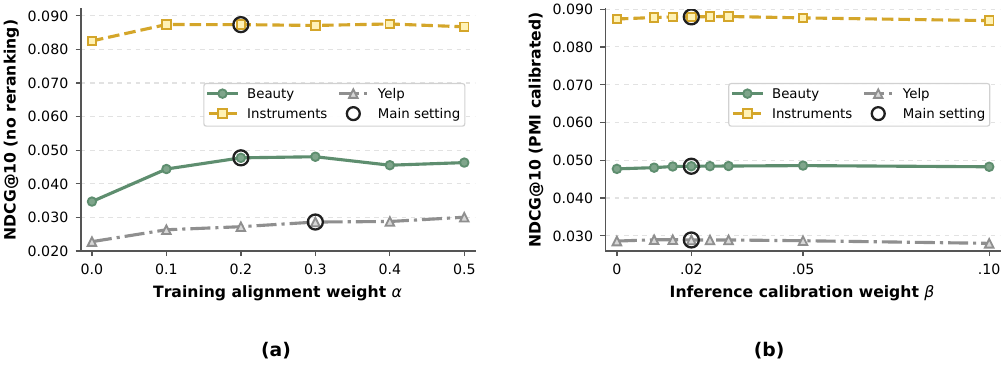}
\vspace{-3mm}
\caption{Sensitivity to alignment and calibration
weights on three datasets.
(a) Uncalibrated NDCG@10 versus $\alpha$.
(b) PMI-calibrated NDCG@10 versus $\beta$.
Outlined markers indicate the main settings,
$\alpha=0.2/0.2/0.3$ and $\beta=0.02$,
rather than the maxima of the test curves.}
\vspace{-3mm}
\label{fig:sensitivity}
\end{figure}

Figure~\ref{fig:sensitivity} reports test NDCG@10 across
the training alignment weight $\alpha$ and inference
calibration weight $\beta$.
The $\alpha$ sweep evaluates uncalibrated models
over the tested grid in $[0,0.5]$.
The highest observed values occur at
$\alpha=0.3$, $0.4$, and $0.5$ for Beauty,
Instruments, and Yelp, respectively.
Yelp is still improving at the upper boundary,
so the sweep does not establish an interior optimum.

The main alignment weights are selected on the
validation set and therefore need not coincide
with the maxima of these test curves.
Outlined markers indicate the main settings
$\alpha=0.2/0.2/0.3$.

For inference calibration, we evaluate the tested
$\beta$ grid over $[0,0.10]$.
The observed maxima occur at $0.05$, $0.03$, and
$0.015$ for Beauty, Instruments, and Yelp,
respectively.
The common setting $\beta=0.02$ is within
0.00017 NDCG@10 of the highest reported value
on each dataset.

\subsection{Monte Carlo Sample Count}
\label{sec:mc_sensitivity}

The conditional term in equation~\eqref{eq:chain_kl} is
estimated using samples from the first-token
target distribution.
We examine $K_{\mathrm{mc}}\in\{2,3,4,5,6\}$
on Instruments while fixing $\alpha=0.2$,
$K_{\mathrm{hist}}=5$, $\gamma=0.7$, and seed 42.
The $K_{\mathrm{mc}}=2$ row reuses the main-experiment
checkpoint; the remaining settings are trained independently.
All settings are evaluated with $\beta=0.02$ and $\tau=30$.

As shown in Table~\ref{tab:mc_sensitivity}, increasing
the sample count does not produce monotonic gains.
NDCG@10 ranges from .0871 to .0891.
The smallest tested count, $K_{\mathrm{mc}}=2$,
achieves .0881, compared with .0891 at
$K_{\mathrm{mc}}=5$.
This sweep supports using a small number of sampled
conditional evaluations, although larger sample
counts can improve the observed result.

\begin{table}[h]
\caption{Monte Carlo sample-count sensitivity on
Instruments. All configurations use $\alpha=0.2$,
$K_{\mathrm{hist}}=5$, $\gamma=0.7$, and seed 42,
with PMI parameters $\beta=0.02$ and $\tau=30$.
Full-SID survival is reported as a percentage.
Bold indicates the best result in each column.}
\label{tab:mc_sensitivity}
\centering
\scriptsize
\setlength{\tabcolsep}{5pt}
\begin{tabular}{cccc}
\toprule
$K_{\mathrm{mc}}$ & H@10 & N@10
& $S_{\mathrm{full}}$ \\
\midrule

2
& .1176 & .0881 & 14.37 \\

3
& .1171 & .0880 & 14.34 \\

4
& .1163 & .0878 & 14.46 \\

5
& \textbf{.1194}
& \textbf{.0891}
& \textbf{14.66} \\

6
& .1156 & .0871 & 14.31 \\

\bottomrule
\end{tabular}
\end{table}

\subsection{History Window and Temporal Decay}
\label{sec:history_sensitivity}

We vary the history window $K_{\mathrm{hist}}$
and decay factor $\gamma$ on Beauty, retraining
each configuration from scratch.
The window sweep fixes $\gamma=0.7$, and the
decay sweep fixes $K_{\mathrm{hist}}=5$.
All runs use $\alpha=0.3$, $K_{\mathrm{mc}}=4$,
$\beta=0.02$, $\tau=30$, and seed 42.

These settings change the history aggregation
used in both training and inference calibration,
so Hit@10 and NDCG@10 measure end-to-end sensitivity.
Full-SID survival is measured before calibration
and reflects the trained model's candidate generation.

\begin{figure*}[t]
\centering
\includegraphics[width=0.96\textwidth]
{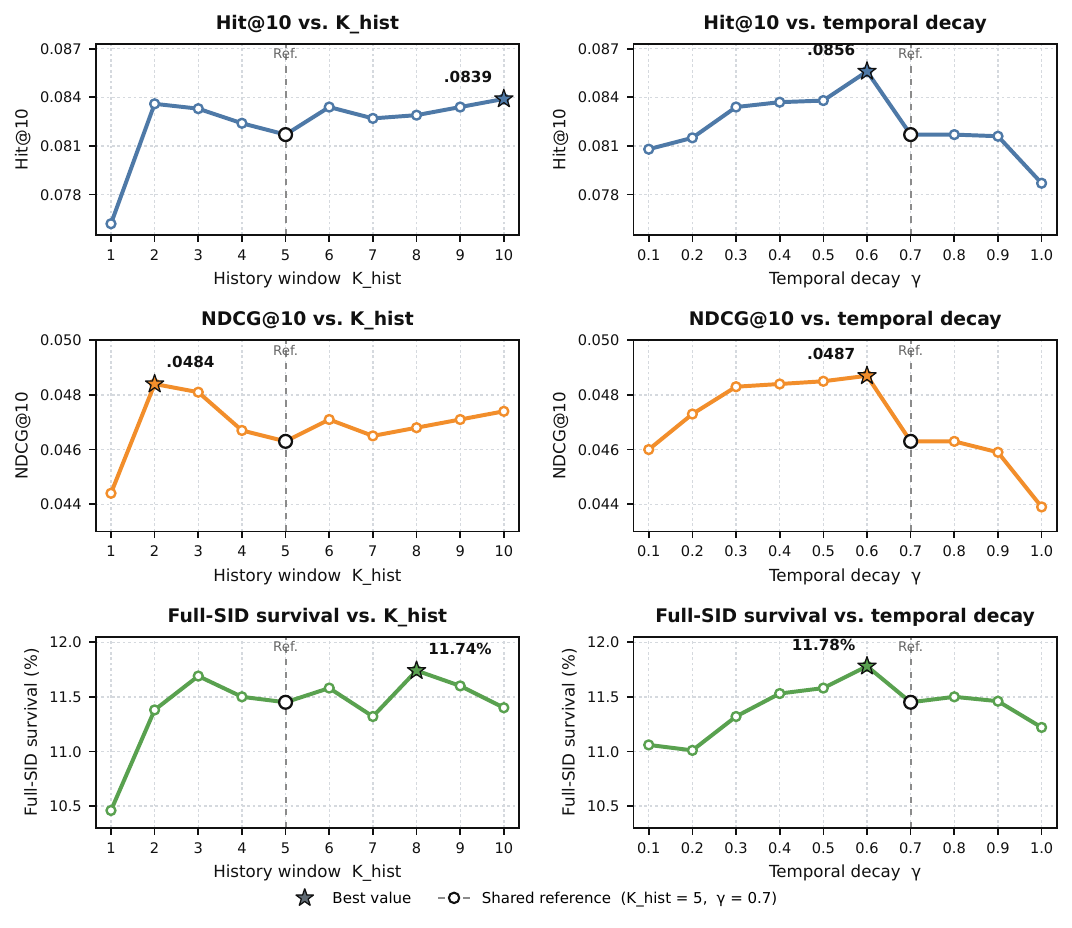}
\caption{History-window and temporal-decay sensitivity
on Beauty. Every configuration is retrained with
$\alpha=0.3$, $K_{\mathrm{mc}}=4$, and seed 42,
then evaluated with $\beta=0.02$ and $\tau=30$.
The left column varies $K_{\mathrm{hist}}$ with
$\gamma=0.7$; the right column varies $\gamma$
with $K_{\mathrm{hist}}=5$.
The rows report Hit@10, NDCG@10, and full-SID
survival, respectively.
Stars indicate the best value in each sweep.
Gray dashed lines and open circles mark the shared
reference configuration
$(K_{\mathrm{hist}},\gamma)=(5,0.7)$.
Full-SID survival is reported as a percentage.}
\label{fig:history_sensitivity}
\end{figure*}

Figure~\ref{fig:history_sensitivity} shows that using
only the most recent interaction gives the lowest
NDCG@10 in the window sweep.
Aggregating at least two interactions improves
the reported metrics, but larger windows do not
provide monotonic gains.
The highest NDCG@10 occurs at
$K_{\mathrm{hist}}=2$, whereas the highest
full-SID survival occurs at $K_{\mathrm{hist}}=8$.

For temporal decay, $\gamma\in[0.3,0.6]$ yields
the highest NDCG@10 values in this configuration.
Uniform weighting ($\gamma=1$) gives .0439,
compared with .0487 at $\gamma=0.6$.
These observations support combining several
recent interactions while discounting older ones
in the tested setting.

\clearpage

\section{Decoder-Only Evaluation}
\label{sec:decoder_only}

\begin{table}[h]
\caption{Decoder-only evaluation with $B=20$
and seed 42. Joint alignment uses $\alpha=0.4$
and $K_{\mathrm{mc}}=2$ across all datasets.
PMI calibration is disabled.
Full-SID survival is reported as a percentage.
Bold indicates the better result within each dataset.}
\label{tab:decoder_only}
\centering
\scriptsize
\setlength{\tabcolsep}{4pt}
\begin{tabular}{llccc}
\toprule
Dataset & Training objective
& H@10 & N@10 & $S_{\mathrm{full}}$ \\
\midrule

\multirow{2}{*}{Beauty}
& NTP
& .0557 & .0302 & 8.36 \\
& Joint alignment
& \textbf{.0605}
& \textbf{.0341}
& \textbf{8.71} \\
\midrule

\multirow{2}{*}{Instruments}
& NTP
& .1004
& \textbf{.0759}
& 12.32 \\
& Joint alignment
& \textbf{.1052}
& .0754
& \textbf{13.02} \\
\midrule

\multirow{2}{*}{Yelp}
& NTP
& .0317 & .0165 & 5.02 \\
& Joint alignment
& \textbf{.0351}
& \textbf{.0188}
& \textbf{5.48} \\
\bottomrule
\end{tabular}
\end{table}

We evaluate whether joint-prefix alignment also
improves target retention in a decoder-only
recommender.
The model is an eight-layer LLaMA-style architecture
with hidden size 128, eight attention heads,
and intermediate size 1024.
It is trained from scratch with effective batch
size 1024, learning rate $5\times10^{-4}$,
bfloat16 precision, and seed 42.

Within each dataset, the NTP and joint-alignment
runs share the data split, SID tokenizer, and
trie-constrained beam search with $B=20$.
Joint alignment uses $\alpha=0.4$ and
$K_{\mathrm{mc}}=2$ on all three datasets.
PMI calibration is disabled.

Table~\ref{tab:decoder_only} shows that joint alignment
improves Hit@10 and full-SID survival on all three
datasets.
NDCG@10 increases from .0302 to .0341 on Beauty
and from .0165 to .0188 on Yelp, but decreases
slightly from .0759 to .0754 on Instruments.
These results support the transfer of the
path-retention benefit to the tested decoder-only
architecture, while showing that improved survival
does not guarantee higher NDCG@10.

\section{Mathematical Details}
\label{sec:math_details}

We derive the joint-KL decomposition and establish
the unbiasedness and sampling variance of the
auxiliary-loss estimator.
The analysis fixes a training example with interaction
history $\mathcal{H}$ and SID-encoded input $X$.
The tokenizer, transition table, and target-construction
hyperparameters are fixed, so the complete target
distribution $q_{12}(\cdot\mid\mathcal{H})$ is
independent of the recommender parameters $\theta$.

\subsection{Chain-Rule Decomposition of Joint KL}
\label{sec:kl_decomposition}

\paragraph{Notation and support.}
For readability, we suppress $\mathcal{H}$ and $X$
and write
\begin{align*}
q_{12}(b_1,b_2)
&=q_{12}(b_1,b_2\mid\mathcal{H}),\\
q_1(b_1)
&=q_1(b_1\mid\mathcal{H}),\\
q_2(b_2\mid b_1)
&=q_2(b_2\mid b_1,\mathcal{H}),\\
p_{\theta,1}(b_1)
&=p_{\theta,1}(b_1\mid X),\\
p_{\theta,2}(b_2\mid b_1)
&=p_{\theta,2}(b_2\mid X,b_1).
\end{align*}
The target $q_{12}$ is a normalized distribution
on the finite vocabulary $C_1\times C_2$, extended
by zero outside $\mathcal{P}_{12}$.

Define its first-token marginal and active
first-token support as
\begin{equation}
q_1(b_1)
=
\sum_{b_2\in C_2}q_{12}(b_1,b_2),
\qquad
\mathcal{A}
=
\{b_1\in C_1:q_1(b_1)>0\}.
\label{eq:appendix_active_support}
\end{equation}
For $b_1\in\mathcal{A}$, the conditional target is
\begin{equation}
q_2(b_2\mid b_1)
=
\frac{q_{12}(b_1,b_2)}{q_1(b_1)},
\qquad
q_{12}(b_1,b_2)
=
q_1(b_1)q_2(b_2\mid b_1).
\label{eq:appendix_target_factorization}
\end{equation}
Branches outside $\mathcal{A}$ have zero joint
target mass and are excluded from the conditional
expectations below.
No conditional target needs to be evaluated on
those branches.

The model distributions are normalized over their
corresponding token vocabularies and satisfy
\begin{equation}
p_{\theta,12}(b_1,b_2)
=
p_{\theta,1}(b_1)\,
p_{\theta,2}(b_2\mid b_1).
\label{eq:appendix_model_factorization}
\end{equation}
If token masks are used, these symbols refer to
the resulting normalized distributions used in
the training objective.

We assume
$p_{\theta,12}(b_1,b_2)>0$
whenever $q_{12}(b_1,b_2)>0$.
On the finite vocabularies, this ensures that
the marginal KL and every conditional KL on
$\mathcal{A}$ are finite.
Terms with zero target probability contribute
zero and are omitted from logarithmic sums.

\paragraph{Proposition 1 (Joint-KL decomposition).}
Under these definitions,
\begin{equation}
\begin{aligned}
\mathrm{KL}(q_{12}\|p_{\theta,12})
&=
\mathrm{KL}(q_1\|p_{\theta,1})\\
&\quad+
\sum_{b_1\in\mathcal{A}}
q_1(b_1)
\mathrm{KL}\!\left(
q_2(\cdot\mid b_1)
\|p_{\theta,2}(\cdot\mid b_1)
\right).
\end{aligned}
\label{eq:appendix_kl_chain}
\end{equation}

\paragraph{Proof.}
Substituting the target and model factorizations
into the definition of KL gives
\begin{equation}
\begin{aligned}
\mathrm{KL}(q_{12}\|p_{\theta,12})
&=
\sum_{b_1\in\mathcal{A}}
\sum_{b_2\in C_2}
q_1(b_1)q_2(b_2\mid b_1)
\log
\frac{q_1(b_1)q_2(b_2\mid b_1)}
{p_{\theta,1}(b_1)p_{\theta,2}(b_2\mid b_1)}\\
&=
\sum_{b_1\in\mathcal{A}}
\sum_{b_2\in C_2}
q_1(b_1)q_2(b_2\mid b_1)
\log\frac{q_1(b_1)}{p_{\theta,1}(b_1)}\\
&\quad+
\sum_{b_1\in\mathcal{A}}
\sum_{b_2\in C_2}
q_1(b_1)q_2(b_2\mid b_1)
\log
\frac{q_2(b_2\mid b_1)}
{p_{\theta,2}(b_2\mid b_1)}.
\end{aligned}
\label{eq:appendix_kl_expansion}
\end{equation}
For each $b_1\in\mathcal{A}$,
$\sum_{b_2\in C_2}q_2(b_2\mid b_1)=1$.
The first term therefore reduces to
\begin{equation}
\sum_{b_1\in\mathcal{A}}
q_1(b_1)
\log\frac{q_1(b_1)}{p_{\theta,1}(b_1)}
=
\mathrm{KL}(q_1\|p_{\theta,1}).
\end{equation}
The second term is
\begin{equation}
\sum_{b_1\in\mathcal{A}}
q_1(b_1)
\mathrm{KL}\!\left(
q_2(\cdot\mid b_1)
\|p_{\theta,2}(\cdot\mid b_1)
\right).
\end{equation}
Combining the two terms establishes
equation~\eqref{eq:appendix_kl_chain}.

\paragraph{Implication for joint alignment.}
The conditional term is nonnegative, so
\begin{equation}
\mathrm{KL}(q_{12}\|p_{\theta,12})
\geq
\mathrm{KL}(q_1\|p_{\theta,1}).
\end{equation}
Even when $q_1=p_{\theta,1}$, mismatched
second-token continuations can leave a positive
joint KL.
Matching the first-token marginal alone therefore
does not ensure that the joint distributions match.
The conditional component supplies the additional
supervision needed to align continuations within
each target-supported branch.
Its effect on finite-beam target survival is
evaluated experimentally in section~\ref{sec:survival}.

\subsection{Unbiased Monte Carlo Estimation}
\label{sec:mc_proof}

\paragraph{Estimator.}
Define the exact marginal term and the conditional
loss associated with an active first-token branch as
\begin{align}
\mathcal{L}_1(\theta)
&=
\mathrm{KL}(q_1\|p_{\theta,1}),\\
f_\theta(b_1)
&=
\mathrm{KL}\!\left(
q_2(\cdot\mid b_1)
\|p_{\theta,2}(\cdot\mid b_1)
\right),
\qquad b_1\in\mathcal{A}.
\end{align}
The chain-rule decomposition gives
\begin{equation}
\mathcal{L}_{12}(\theta)
=
\mathcal{L}_1(\theta)
+
\sum_{b_1\in\mathcal{A}}
q_1(b_1)f_\theta(b_1).
\label{eq:appendix_exact_objective}
\end{equation}

For an integer $K_{\mathrm{mc}}\geq1$, draw
independent samples with replacement:
\begin{equation}
b_1^{(1)},\ldots,b_1^{(K_{\mathrm{mc}})}
\overset{\mathrm{i.i.d.}}{\sim}q_1.
\end{equation}
All sampled branches belong to $\mathcal{A}$
with probability one.
The estimator is
\begin{equation}
\widetilde{\mathcal{L}}_{12}(\theta)
=
\mathcal{L}_1(\theta)
+
\frac{1}{K_{\mathrm{mc}}}
\sum_{k=1}^{K_{\mathrm{mc}}}
f_\theta(b_1^{(k)}).
\label{eq:appendix_mc_estimator}
\end{equation}
The marginal term is computed exactly.
For each sampled branch, the conditional KL is
also computed exactly over the second-token
vocabulary.
Sampling approximates only the expectation over
first-token branches.

All expectations and variances below are taken
over these branch samples, with the training
example and model distributions held fixed.

\paragraph{Proposition 2 (Unbiased objective and gradient).}
The estimator satisfies
\begin{equation}
\mathbb{E}
\left[
\widetilde{\mathcal{L}}_{12}(\theta)
\right]
=
\mathcal{L}_{12}(\theta).
\label{eq:appendix_loss_unbiased}
\end{equation}
If the model probabilities on the target support
are differentiable and positive in a neighborhood
of the parameter value under consideration, then
\begin{equation}
\mathbb{E}
\left[
\nabla_\theta
\widetilde{\mathcal{L}}_{12}(\theta)
\right]
=
\nabla_\theta\mathcal{L}_{12}(\theta).
\label{eq:appendix_gradient_unbiased}
\end{equation}

\paragraph{Proof of objective unbiasedness.}
Each sampled branch has distribution $q_1$.
Consequently,
\begin{equation}
\mathbb{E}
\left[
f_\theta(b_1^{(k)})
\right]
=
\sum_{b_1\in\mathcal{A}}
q_1(b_1)f_\theta(b_1).
\end{equation}
By linearity of expectation,
\begin{equation}
\begin{aligned}
\mathbb{E}
\left[
\widetilde{\mathcal{L}}_{12}(\theta)
\right]
&=
\mathcal{L}_1(\theta)
+
\frac{1}{K_{\mathrm{mc}}}
\sum_{k=1}^{K_{\mathrm{mc}}}
\mathbb{E}
\left[
f_\theta(b_1^{(k)})
\right]\\
&=
\mathcal{L}_1(\theta)
+
\sum_{b_1\in\mathcal{A}}
q_1(b_1)f_\theta(b_1)\\
&=
\mathcal{L}_{12}(\theta).
\end{aligned}
\end{equation}

\paragraph{Proof of gradient unbiasedness.}
The entire target distribution is independent
of $\theta$, so $q_1$, $q_2$, and $\mathcal{A}$
remain fixed during differentiation.
Because $\mathcal{A}$ is finite and each branch
loss is differentiable,
\begin{equation}
\nabla_\theta\mathcal{L}_{12}(\theta)
=
\nabla_\theta\mathcal{L}_1(\theta)
+
\sum_{b_1\in\mathcal{A}}
q_1(b_1)\nabla_\theta f_\theta(b_1).
\label{eq:appendix_exact_gradient}
\end{equation}
Treating sampled token indices as constants
during differentiation gives
\begin{equation}
\nabla_\theta\widetilde{\mathcal{L}}_{12}(\theta)
=
\nabla_\theta\mathcal{L}_1(\theta)
+
\frac{1}{K_{\mathrm{mc}}}
\sum_{k=1}^{K_{\mathrm{mc}}}
\nabla_\theta f_\theta(b_1^{(k)}).
\end{equation}
Taking expectations,
\begin{equation}
\begin{aligned}
\mathbb{E}
\left[
\nabla_\theta\widetilde{\mathcal{L}}_{12}(\theta)
\right]
&=
\nabla_\theta\mathcal{L}_1(\theta)
+
\frac{1}{K_{\mathrm{mc}}}
\sum_{k=1}^{K_{\mathrm{mc}}}
\mathbb{E}
\left[
\nabla_\theta f_\theta(b_1^{(k)})
\right]\\
&=
\nabla_\theta\mathcal{L}_1(\theta)
+
\sum_{b_1\in\mathcal{A}}
q_1(b_1)\nabla_\theta f_\theta(b_1)\\
&=
\nabla_\theta\mathcal{L}_{12}(\theta).
\end{aligned}
\end{equation}
No derivative of the sampling distribution is
required because it is independent of $\theta$.

The result holds for every $K_{\mathrm{mc}}\geq1$.
Choosing $K_{\mathrm{mc}}=2$ changes computational
cost and sampling variance, but not the expected
auxiliary objective or gradient.

\paragraph{Combined training objective.}
For a fixed alignment weight $\alpha$, define
the exact and sampled training objectives as
\begin{align}
\mathcal{J}(\theta)
&=
\mathcal{L}_{\mathrm{NTP}}(\theta)
+
\alpha\mathcal{L}_{12}(\theta),\\
\widetilde{\mathcal{J}}(\theta)
&=
\mathcal{L}_{\mathrm{NTP}}(\theta)
+
\alpha\widetilde{\mathcal{L}}_{12}(\theta).
\end{align}
The NTP term is unaffected by branch sampling.
Therefore,
\begin{equation}
\mathbb{E}
\left[
\widetilde{\mathcal{J}}(\theta)
\right]
=
\mathcal{J}(\theta),
\qquad
\mathbb{E}
\left[
\nabla_\theta\widetilde{\mathcal{J}}(\theta)
\right]
=
\nabla_\theta\mathcal{J}(\theta),
\label{eq:appendix_total_gradient}
\end{equation}
where the gradient identity assumes that
$\mathcal{L}_{\mathrm{NTP}}$ is differentiable.

\paragraph{Sampling variance.}
For fixed $\theta$ and a fixed training example,
let
\begin{equation}
\mu_f
=
\mathbb{E}_{b_1\sim q_1}[f_\theta(b_1)],
\qquad
\sigma_f^2
=
\mathbb{E}_{b_1\sim q_1}
\left[
(f_\theta(b_1)-\mu_f)^2
\right].
\end{equation}
These moments are finite because $\mathcal{A}$
is finite and each branch loss is finite.

The marginal term contributes no branch-sampling
variance.
Independence of the sampled branches eliminates
cross-sample covariance terms, giving
\begin{equation}
\begin{aligned}
\operatorname{Var}
\left[
\widetilde{\mathcal{L}}_{12}
\right]
&=
\operatorname{Var}
\left[
\frac{1}{K_{\mathrm{mc}}}
\sum_{k=1}^{K_{\mathrm{mc}}}
f_\theta(b_1^{(k)})
\right]\\
&=
\frac{1}{K_{\mathrm{mc}}^2}
\sum_{k=1}^{K_{\mathrm{mc}}}
\operatorname{Var}
\left[
f_\theta(b_1^{(k)})
\right]\\
&=
\frac{\sigma_f^2}{K_{\mathrm{mc}}}.
\end{aligned}
\label{eq:appendix_mc_variance}
\end{equation}

The same reasoning applies to the gradient.
Let
\begin{equation}
g_\theta(b_1)=\nabla_\theta f_\theta(b_1),
\qquad
\Sigma_g
=
\operatorname{Cov}_{b_1\sim q_1}
[g_\theta(b_1)].
\end{equation}
The covariance exists under the differentiability
and positivity assumptions above because the
active support is finite.
Since the marginal gradient is deterministic
with respect to branch sampling,
\begin{equation}
\begin{aligned}
\operatorname{Cov}
\left[
\nabla_\theta\widetilde{\mathcal{L}}_{12}
\right]
&=
\operatorname{Cov}
\left[
\frac{1}{K_{\mathrm{mc}}}
\sum_{k=1}^{K_{\mathrm{mc}}}
g_\theta(b_1^{(k)})
\right]\\
&=
\frac{\Sigma_g}{K_{\mathrm{mc}}}.
\end{aligned}
\label{eq:appendix_gradient_variance}
\end{equation}
For the combined objective,
\begin{equation}
\operatorname{Cov}
\left[
\nabla_\theta\widetilde{\mathcal{J}}
\right]
=
\frac{\alpha^2\Sigma_g}{K_{\mathrm{mc}}}.
\end{equation}

These identities quantify branch-sampling noise
at fixed model parameters.
They do not include other sources of training
randomness, such as minibatch selection or dropout.
Increasing $K_{\mathrm{mc}}$ reduces this component
of estimator variance, but does not imply monotonic
improvements in final recommendation metrics,
which also depend on the optimization trajectory.

\end{document}